\documentclass[12pt]{article}
\usepackage{amssymb}
\usepackage{amsmath}
\usepackage{hyperref}
\usepackage[dvips]{graphicx}
\usepackage{bm}
\usepackage{latexsym}
\begin{document}
\title{\bf   AdS black brane coupled to  non-abelian logarithmic gauge theory  and  color DC conductivity }
\author{ Mehdi Sadeghi\thanks{Corresponding author: Email:  mehdi.sadeghi@abru.ac.ir}\hspace{2mm}\\
{\small {\em Department of Physics, School of Sciences,}}\\
        {\small {\em Ayatollah Boroujerdi University, Boroujerd, Iran}}
       }
\date{\today}
\maketitle

\abstract{ In this paper, we consider Einstein-Hilbert gravity in the presence of cosmological constant and non-abelian nonlinear electromagnetic field of logarithmic type which is minimally coupled to gravity. First, black brane solution of this model is introduced  and then color non-abelian DC conductivity is calculated for this solution by using AdS/CFT duality. Our result recovers the Yang-Mills model in $\lambda \to 0$ limit.}\\



\section{Introduction} \label{intro}

Maxwell’s electromagnetic field theory that is a linear theory, suffers from some problems \cite{Delphenich:2003yw,Delphenich:2006ec,EslamPanah:2021xaf}, singularity in the origin of particle, infinite self-energy, description of the self-interaction of virtual electron-positron pairs \cite{W. Heisenberg_1936}, the radiation propagation inside specific materials \cite{DeLorenci:2001gf}. Heisenberg and Euler have shown that quantum electrodynamics (QED) due to loop corrections gives nonlinear electrodynamics (NED) \cite{W. Heisenberg_1936}.
NED's coupled to general relativity (GR) can explain inflation and current acceleration of the universe \cite{Camara:2004ap}-\cite{Garcia-Salcedo:2013cxa}. Logarithmic gauge theory \cite{Gaete:2013dta},\cite{Dehghani:2019xjc},\cite{Dehghani:2021civ},\cite{Dehghani:2022xac},\cite{Dehghani:2021cyb}, arcsin-electrodynamics \cite{Kruglov_2015}, Born-Infeld theory \cite{Born1,Born2} have been introduced as nonlinear electrodynamics (NED) theories to solve some of problems in Maxwell’s theory. The NED field gives more information in higher magnetized neutron stars and pulsars \cite{Bialynicka-Birula}\cite{H. J. Mosquera} and this NED theories remove both of the big bang and black hole singularities \cite{Corda:2009xd}-\cite{AyonBeato:1999ec} by modifying space-time geometry.\\
There are three kinds of Born-Infeld (BI) actions, Born-Infeld nonlinear electromagnetic (BINEF), logarithmic form of nonlinear electromagnetic field (LNEF) and exponential form of nonlinear electromagnetic field (ENEF). The color DC
conductivity bound of Einstein–Born–Infeld AdS black brane was studied in \cite{Sadeghi:2021qou} and the same calculation for logarithmic form of Born-Infeld (BI) action is investigated in this paper. We also will consider the exponential form of nonlinear electromagnetic field in our future work.\\
We want to study the field theory dual of the logarithmic gauge theory by AdS/CFT duality \cite{Maldacena:1997re}. Therefore, we should calculate the transport coefficients like conductivity and shear viscosity to entropy density.
Hydrodynamics is an effective theory for describing near-equilibrium phenomena at late
time and large distance \cite{Landau,Son,Policastro2001}. Conservation equations together with constitutive relations in a gradient expansion around equilibrium is constructed hydrodynamics equations. These equations up to first order of derivative expansion are as follows, \\
\begin{align}
& \nabla _{\mu } J^{\mu } =0\,\,\,\,\,,\,\,\nabla _{\mu } T^{\mu \nu} =0,\\
& J^{\mu } =n \, u^{\mu }-\sigma T P^{\mu \nu }\partial_{\nu}(\frac{\mu}{T}),\nonumber\\
& T^{\mu \nu } =(\rho +p)u^{\mu } u^{\nu } +pg^{\mu \nu } -\sigma ^{\mu \nu },\nonumber\\
&\sigma ^{\mu \nu } = {P^{\mu \alpha } P^{\nu \beta } } [\eta(\nabla _{\alpha } u_{\beta } +\nabla _{\beta } u_{\alpha })+ (\zeta-\frac{2}{3}\eta) g_{\alpha \beta } \nabla .u],\nonumber\\& P^{\mu \nu }=g^{\mu \nu}+u^{\mu}u^{\nu}, \nonumber
\end{align}
 where $n$, $\eta$, $\zeta $, $\sigma ^{\mu \nu }$ , $\sigma$ and $P^{\mu \nu }$ are charge density, shear viscosity, bulk viscosity, shear tensor, conductivity and projection operator, respectively \cite{Kovtun2012}. $\eta$, $\zeta $ and $\sigma $ are known as transport coefficients.

The fluid-gravity correspondence as a subclass of gauge-gravity duality states that
black holes in AdS space-time are dual to stationary solutions of the equations of relativistic hydrodynamics \cite{Son}-\cite{J.Bhattacharya2014}.  Conductivity is calculated by Green-Kubo formula and Witten prescription.\\ 
\begin{equation} \label{kubo}
\sigma^{ij} _{ab}(k_{\mu})=-\mathop{\lim }\limits_{\omega \to 0} \frac{1}{\omega } \Im G^{ij}_{ab}(k_{\mu}),
\end{equation} 
where $a,b$ and $i,j$ indices refer to  $SU(2)$ group symmetry and translational symmetry in spatial directions of $x,y$, respectively. There is no difference between upper and lower indices on the conformal field theory (CFT) side because the metric of the boundary theory is flat spacetime with Minkowski metric. There is a lower bound for DC conductivity as $\sigma \geq \frac{1}{e^2}=1$ where $e$ is the charge of the gauge field but $e$ is not the unit of charge in the boundary theory \cite{Grozdanov:2015qia,Grozdanov:2015djs}. This bound is also saturated in graphene\cite{Ziegler}. This bound is violated for massive gravity \cite{Baggioli:2016oqk}, models with abelian case \cite{Baggioli:2016oju} and non-abelian Born-Infeld theory\cite{Sadeghi:2021qou}. In this paper, we want to investigate the effects of non-abelian logarithmic gauge theory as a nonlinear electrodynamics gauge field on the conductivity bound of strongly coupled dual theory.
\section{  AdS black brane coupled to non-abelian logarithmic gauge theory Solution}
\label{sec2}

\indent The 4-dimensional action of non-abelian logarithmic gauge theory with negative cosmological constant is \cite{Soleng:1995kn},\cite{Chaloshtary:2019qvv},
\begin{eqnarray}\label{action}
S=\frac{1}{16\pi G}\int d^{4}  x\sqrt{-g} \bigg[R-2\Lambda -\frac{1}{\lambda}
\log\big(1+\lambda \mathcal{F}\big)\bigg],
\end{eqnarray}
where $R$ is the Ricci scalar, $l$ the AdS radius, $\mathcal{F}={\bf{Tr}}( F_{\mu \nu }^{(a)} F^{(a)\, \, \mu \nu })$ is Yang-Mills invariant and $\lambda$ is the value of the non linear coupling constant parameter. The trace element stands for ${\bf{Tr}}(.)=\sum_{i=1}^{3}(.).$ \\
$F_{\mu \nu }^{(a)}$ is the $SU(2) $ Yang-Mills  field strength tensor,
\begin{align} \label{YM}
F_{\mu \nu } =\partial _{\mu } A_{\nu } -\partial _{\nu } A_{\mu } -i[A_{\mu }, A_{\nu }],
\end{align}
in which the gauge coupling constant is 1, $A_{\nu }$'s are the $SU(2)$ gauge group Yang-Mills potentials. When $\lambda \to 0 $ the logarithmic term gets transformed into standard linear non-abelian Yang-Mills term \cite{Shepherd:2015dse}.\\
We get to the equations of motion by varying the action \ref{action} with respect to  $g_{\mu \nu } $ and $A^{(a)}_{\nu}$,
\begin{equation}\label{EH}
R_{\mu \nu }-  \tfrac{1}{2}  g_{\mu \nu }R+\Lambda g_{\mu \nu }+  \tfrac{1}{2\lambda} g_{\mu \nu } \log(1 + \lambda \mathcal{F}) -  \frac{2  F^{(a)}_{\mu}\, ^{\alpha } F^{(a)}_{\nu \alpha } }{1 + \
	\lambda \mathcal{F}}=0,
\end{equation}
\begin{equation}\label{EOM2}
\nabla_{\mu}\bigg(\frac{ F^{{(a)}\mu \nu }}{1 + \lambda \mathcal{F}}\bigg)=0.
\end{equation}
Our goal is to find asymptotically AdS black brane solution in four dimensional flat symmetric
spacetime, so we consider the following ansatz,
\begin{equation}\label{metric1}
ds^{2} =-f(r)dt^{2} +\frac{dr^{2} }{f(r)} +\frac{r^2}{l^2}(dx^2+dy^2),
\end{equation}
where $f(r)$ is the metric function that should be determined and $l$ is a constant relating to the cosmological constant.\\
We write the gauge field in terms of matrix $\sigma_3= \begin{pmatrix}1 & 0 \\ 0 & -1\end{pmatrix}$, which is the generators of $SU(2)$ \cite{Shepherd:2015dse}.\\\\
\begin{equation}\label{background}
{\bf{A}}^{(a)} =h(r)dt\begin{pmatrix}1 & 0 \\ 0 & -1\end{pmatrix}.
\end{equation}
Eq.\ref{EOM2} can be written as below by considering \ref{EOM2} and \ref{metric1},
\begin{equation}
2 h' \Big(\lambda  h' \left(r h''-2 h'\right)+1\Big)+r h''=0.
\end{equation}
One could solve the above differential equation to obtain the function $h(r)$ as following,

\begin{equation}\label{h}
h(r)=\mu-\frac{4 q^2 }{3 r }\,_2F_1\left[\frac{1}{2},\frac{1}{4};\frac{5}{4};-\frac{8 \lambda q^{2} }{r^4}\right] +\frac{r^3}{12 q \lambda}  \left(\sqrt{1+\frac{8\lambda q^2}{r^4}}-1\right),
\end{equation}
$\mu$ is a constant of integration and called chemical potential of quantum field that locates on the boundary of AdS spacetime. It is found by applying the regularity condition on the horizon i.e. $A_t(r_h) = 0$ \cite{Dey:2015poa}.\\
\begin{equation}
\mu=\frac{4 q^2 }{3 r_h }\,_2F_1\left[\frac{1}{2},\frac{1}{4};\frac{5}{4};-\frac{8 \lambda q^{2} }{r_h^4}\right] -\frac{r_h^3}{12 q \lambda}  \left(\sqrt{1+\frac{8\lambda q^2}{r_h^4}}-1\right),
\end{equation}
for the $\lambda=0$, $h(r)$ will be as,
\begin{align}
h(r)=\frac{q}{r},
\end{align}
\noindent where $q$ is integration constant which is related to the electric charge. The invariant scalar $\mathcal{F}_{YM}={\bf{Tr}}( F_{\mu \nu }^{(a)} F^{(a)\, \, \mu \nu })$ for  fields is,
\begin{equation}
  F_{tr}^{(a)} =-F_{rt}^{(a)}=h'(r)\begin{pmatrix}1 & 0 \\ 0 & -1\end{pmatrix}=\frac{2qr^{-2}}{1+\sqrt{1+\frac{8\lambda q^2}{r^4}}}\begin{pmatrix}1 & 0 \\ 0 & -1\end{pmatrix},
\end{equation}
by setting $\lambda=0$, our results decrease to non-abelian Yang-Mills solution.\\
The Bianchi identity is satisfied,\\
\begin{equation}
\nabla_{\alpha }F^{(a)}_{\mu \nu}+\nabla_{\nu }F^{(a)}_{\alpha \mu}+\nabla_{\mu }F^{(a)}_{ \nu \alpha}=0.
\end{equation}
By considering $tt$ component of  Eq. (\ref{EH}), 
\begin{equation}\label{eom}
rf'(r)+f(r)+\Lambda r^2+\frac{r^2}{2\lambda}\log \left(1-4 \lambda h'^2\right)+\frac{2 r^2 h'^2}{1-4 \lambda h'^2} =0,
\end{equation} 
where $1-4 \lambda h'^2>0$.\\
$f(r)$ is obtained by solving Eq.(\ref{eom})
\begin{align}\label{sol}
&f(r)=\frac{m}{r}-\frac{\Lambda r^2}{3}+\frac{16 q^2}{9r^2} \, _2F_1\left[\frac{1}{2},\frac{1}{4};\frac{5}{4};-\frac{ 8\lambda q^2 }{r^4}\right]\nonumber\\&+\frac{r^2}{6\lambda}\Bigg[\log\Big(\frac{1+\sqrt{1+\frac{8\lambda q^2}{r^4}}}{2}\Big)+\frac{5}{3}\Bigg(1-\sqrt{1+\frac{8\lambda q^2}{r^4}}\Bigg)\Bigg],
\end{align}
where $m$ is an integration constant and $_2F_1$ is the hypergeometric function. Notice $r$ is the radial coordinate that put us from bulk to the boundary.\\
We notice that the topology of event horizon in our model is flat so the extrinsic curvature is zero, $\kappa=0$. But, the solution of Einstein's equations in \cite{Hendi:2013dwa} was considered black hole so the extrinsic curvature is 1, $\kappa=1$. \\
Hawking temperature for this black brane is,
  \begin{align}\label{Temp}
  T &=\frac{f'(r_h)}{4 \pi}=\frac{-1}{4 \pi }\Bigg(\frac{m}{r_h^2}+\frac{2\Lambda r_h}{3}+\frac{16 q^2}{9 \pi  r_h^3} \, _2F_1\left(\frac{1}{4},\frac{1}{2};\frac{5}{4};\frac{8 q^2 \lambda }{r_h^4}\right) +\frac{16 q^2}{9 r_h^3 \sqrt{1-\frac{8 \lambda  q^2}{r_h^4}}}\nonumber\\&+\frac{\bigg(2 \sqrt{\frac{8 \lambda  q^2}{r_h^4}+1}-2+\log8\bigg)r_h^8+8 \lambda  q^2 r_h^4 (\log8-2)}{9 \lambda  r_h^3 \left(8 \lambda  q^2+r_h^4\right)}
  +\frac{3 r_h^4}{\lambda } \log \left(1+\sqrt{1+\frac{8 \lambda  q^2}{r_h^4}}\right)
  \Bigg).
  \end{align}
Event horizon is located on where  $f(r_h)=0$ and $m$ is identified by this condition.

\section{Color non-abelian DC conductivity}
\label{sec3}
\indent For calculating retarded Green's function $ <J^{\mu}(\omega)J^{\nu}(-\omega)>$, we perturb the gauge field as $A_{\mu} \to A_{\mu}+\tilde{A}_{\mu}$ and expand the action up to second order of perturbed part, $\tilde{A}_{\mu}$ \cite{Baggioli:2016oju}- \cite{Policastro2002}. By reading $J_a^{\mu}$ from the perurbed part of the action as,
 \begin{equation}
 S \to S +\int d^4x\,\, A^a_{\mu}J_a^{\mu}.
 \end{equation}
Retarded Green’s function read from AdS/CFT correspondence as follows,\\
\begin{equation}
\sigma^{\mu \nu}_{ab}(\omega)=\frac{1}{i \omega}<J^{\mu}_{a}(\omega)J^{\nu}_{b}(-\omega)>=\frac{\delta^2 S}{\delta \tilde{A}^{0\,(a)}_{\mu} \delta \tilde{A}^{0(b)}_{\nu}},
\end{equation}
where $\tilde{A}^0_{\nu}$ is the value of gauge field on the boundary. We can calculate non-abelian color  DC conductivity by using Green-Kubo formula,
\begin{equation} \label{kubo2}
\sigma^{ij}_{ab} (k_{\mu})=-\mathop{\lim }\limits_{\omega \to 0} \frac{1}{\omega } \Im G^{ij}_{ab}(k_{\mu}),
\end{equation}
So DC conductivities are related to the retarded Green's function of the boundary current $\mathop{J}^i$.
The conductivities along $i$ and $j$ directions can be expressed as $\sigma^{ij}_{ab}(x,y)=\frac{\delta J^i_a(x)}{\delta E^b_j(y)}$ respectively. There is a rotation symmetry in $xy$ plane so color DC conductivity is a scalar quantity,
\begin{equation}
\sigma^{ij}_{ab}=\sigma_{ab}\delta^{ij}.
\end{equation}
 Now, we consider the perturbation part of  gauge  field as $\tilde{A}_x=\tilde{A}_x(r)e^{-i\omega t}$ in which $\omega$ is small since hydrodynamics regime is valid.  By inserting the perturbed part of gauge field into action Eq.(\ref{action}) and keeping terms up to second order of $\tilde{A}$ we have,\\
\begin{align}\label{action-2}
S^{(2)}&=\int d^4x \frac{4}{ f\left(1-4\lambda h'^2\right)} \left[f^2 \left((\partial_r\tilde{A}_x^{(1)})^2+(\partial_r\tilde{A}_x^{(2)})^2+(\partial_r\tilde{A}_x^{(3)})^2\right)\right. \nonumber\\
&\left.-\Big((\tilde{A}_x^{(1)})^2+(\tilde{A}_x^{(2)})^2+\tilde{A}_x^{(3)})^2\Big)\omega ^2 + 4\Big(\tilde{A}_x^{(1)})^2+\tilde{A}_x^{(2)}\Big)h ^2\right].
\end{align}

By variation of the action $S^{(2)}$ with respect to $\tilde{A}_x ^a$ we have,

\begin{equation}\label{PerA1}
f\left(f
\tilde{A}_x^{(1)'}\right)'+(\omega ^2-4h^2)\tilde{A}_x^{(1)}+\frac{8 \lambda f^2 \tilde{A}_x^{(1)'} h' h''}{1-4\lambda h'^2 }=0,
\end{equation}
\begin{equation}\label{PerA2}
f\left(f
\tilde{A}_x^{(2)'}\right)'+(\omega ^2-4 h^2)\tilde{A}_x^{(2)}+\frac{8 \lambda f^2 \tilde{A}_x^{(2)'} h' h''}{1-4\lambda h'^2}=0,
\end{equation}

\begin{equation}\label{PerA3}
 f \left(f
\tilde{A}_x^{(3)'}\right)'+\omega ^2\tilde{A}_x^{(3)}
+\frac{ 8 \lambda f^2 \tilde{A}_x^{(3)'} h'
h''}{1-4\lambda h'^2}=0.     
\end{equation}
Firstly, we solve Eq.(\ref{PerA1}), Eq.(\ref{PerA2}) and Eq.(\ref{PerA3}) near the horizon. Therefore, by expanding theses equations near event horizon and considering the relation $f\sim4\pi T(r-r_h)$ from definition of Hawking temperature we find the solution of $\tilde{A}_x^{(a)}$ as follows,
\begin{align}
\tilde{A}_x^{(a)}\sim (r-r_h)^{z_a} \, , \qquad a=1,2,3\,\,\,\,,   
\end{align}
where,
\begin{align}\label{z12}
z_1&=z_2=\pm i \frac{\sqrt{h(r_h)^2+\omega ^2}}{4 \pi T} , \\
\label{z3}
z_3&=\pm i \frac{\omega }{4 \pi T}.
\end{align}
To solve Eq.(\ref{PerA1}), Eq.(\ref{PerA2}) and Eq.(\ref{PerA3}), we apply these following ansatzs,
\begin{align}\label{EOMA1}
\tilde{A}_x^{(1)}=\tilde{A}^{(1)}_{\infty}\Big(\frac{21 \lambda}{r^2 (2-7 \lambda \Lambda )}f\Big)^{z_1}\Big(1+i\omega h_1(r)+\cdots\Big) ,
\end{align}
\begin{align}\label{EOMA2}
\tilde{A}_x^{(2)}=\tilde{A}^{(2)}_{\infty}\Big(\frac{21 \lambda}{r^2 (2-7 \lambda \Lambda )}f\Big)^{z_2}\Big(1+i\omega h_2(r)+\cdots\Big) ,
\end{align}
\begin{align}\label{EOMA3}
\tilde{A}_x^{(3)}=\tilde{A}^{(3)}_{\infty}\Big(\frac{21 \lambda}{r^2 (2-7 \lambda \Lambda )}f\Big)^{z_3}\Big(1+i\omega h_3(r)+\cdots\Big) ,
\end{align}
where $\tilde{A}^{(a)}_{\infty}$ is the value of fields in the boundary and $z_i$'s are the minus sign of \ref{z12} and \ref{z3}.\\
By inserting Eq.\ref{EOMA1}     into   Eq.(\ref{PerA1})  and considering to the first order of $\omega$, we obtain,
\begin{align}\label{h1}
&-4 r^2 h^2 \left(4 \mathit{k} h'^2-1\right) \left(4 \pi  T h_1-\log \left(-\frac{21 \mathit{k} f}{r^2 (7
	\mathit{k} \Lambda -2)}\right)\right)\nonumber\\& +r f \Bigg(\left(4 \mathit{k} h'^2-1\right) \bigg(f' \left(4 \pi  r T h_1'+2\right)-r f''\bigg)+8 \mathit{k} r
f' h' h''\Bigg)\nonumber\\&+
2 f^2 \Bigg(2 \pi  r^2 T h_1''\left(4 \mathit{k} h'^2-1\right)-4 \mathit{k} h'
\bigg(2 r h'' \left(2 \pi  r T h_1'+1\right)+h'\bigg)+1\Bigg)=0.
\end{align}


The equation for $h_2(r)$ is the same as $h_1(r)$.\\
By inserting Eq.\ref{EOMA3}     into   Eq.(\ref{PerA3})  and considering to the first order of $\omega$, we obtain,
\begin{align}
& \frac{1}{r} \Bigg(\left(4\lambda h'^2-1\right) \bigg(f' \left(4 \pi  T r h_3'+2\right)-r f''\bigg)+8 \lambda r f'h'
h''\Bigg)\nonumber\\ & +\frac{2}{r^2}f \Bigg(1-4\lambda h' \bigg(2 r h'' \left(2 \pi  r T h_3'+1\right)+h'\bigg)+2 \pi  r^2 T h_3'' \left(1-4\lambda h'^2\right)\Bigg)=0.
\end{align}
The solution of $h_3(r)$ is as,
\begin{align}
& h_3(r) =C_3 + \int_1^r \Bigg(\frac{uf'(u)-2 f(u)}{4 \pi  T u f(u)}+\frac{  C_4  \left(1-4\lambda h'(u)^2\right)}{  f(u)}\Bigg)du,
\end{align}
where $C_3$ and $C_4$ are integration constants. The behavior of $h_3(r)$ near horizon is,
\begin{equation}\label{C4}
h_3 \approx  \bigg(\frac{C_4 (1-4\lambda h'(r_h)^2) +1 }{4 \pi  T}\bigg) \log(r-r_h)+\text{finite terms}.
\end{equation}
 $C_4$ is determined by applying regularity of $h_3(r)$ at horizon as,
\begin{equation}
C_4=\frac{1 }{4\lambda h'(r_h)^2-1} .
\end{equation}
By plugging the solution of $\tilde{A}_x^{(3)}$ in Eq.(\ref{action-2}) and variation with respect to $\tilde{A}^{(3)}_{\infty}$ , Green's function  can be read as,
\begin{equation} \label{Green1}
 G_{xx}^{(33)} (\omega ,\vec{0})=\frac{(\tilde{A}_x^{(3)})^{*}f(r) \partial _{r}\tilde{A}_{x}^{(3)}}{4\lambda h'^2-1}\bigg|_{r \to \infty}=i \omega C_4,
\end{equation}
by substituting $f(r)$ , $h_3(r)$ , $h'(r)$ in above equation and using Eq.(\ref{kubo2}) we have,  
\begin{eqnarray}\label{sigma33}
\sigma_{xx}^{(33)}=-\mathop{\lim }\limits_{\omega \to 0} \frac{1}{\omega } \Im G^{ij}(k_{\mu}) =\frac{1}{ 1-4\lambda h'^2(r_h) },
\end{eqnarray}
The conductivity bound is violated for $ 1-4\lambda h'^2(r_h) > 1 $  and it is preserved for $ 0<1-4\lambda h'^2(r_h) \leq 1 $ .\\
In the limit of $\lambda \to 0$  we have non-abelian Yang-Mills theory and the conductivity bound is saturated.
\begin{eqnarray}
\sigma_{xx}^{(33)} =1.
\end{eqnarray}
The value of $\sigma_{xx}^{(11)}$ and  $\sigma_{xx}^{(22)}$ is calculated by the same procedure of $\sigma_{xx}^{(33)}$ as following,
\begin{eqnarray}
\sigma_{xx}^{(11)} =\sigma_{xx}^{(22)} =0.
\end{eqnarray}
It means that the color non-abelian DC conductivity in terms of color indices is diagonal and also Ohm's law is diagonal in this model. If we consider the gauge field \ref{background} in  $\sigma_{1}$ or $\sigma_{2}$ directions, conductivity will be non-zero in these directions.

 \section{Conclusion}

\noindent We introduced Einstein  non-abelian logarithmic gauge theory AdS black brane solution and calculated the non-abelian  color DC conductivity for this model. There is a conjecture that conductivity is bounded by the universal value $\sigma \geq 1$ {\footnote{ We consider $\frac{1}{e^2}=1$}. Our outcome shows that the conductivity bound is violated for $ 1-4\lambda h'^2(r_h) > 1 $  and preserved for $0<1-4\lambda h'^2(r_h) \leq 1$  in non-abelian logarithmic gauge theory but this bound is saturated for Yang-Mills theory.\cite{S Parvizi}. The conductivity bound can be violated in nonlinear model, depending on the form of the nonlinear theory, or coupling of $F^2$ to the scalar field. Our interpretation is that nonlinearities are somehow similar to electron-electron couplings and therefore the violation related to Mott insulators. It can be concluded that the conductivity is the derivative of the Lagrangian with respect to $F^2$ computed at the event horizon and it is saturated for the Yang-Mills theory but it can be violated for the nonlinear model.\\
The ratio of shear viscosity to entropy density for this model is $\frac{\eta }{s}=\frac{1 }{4 \pi}$. Therefore, the Kovtun-Son-Starinet (KSS) bound \cite{Kovtun:2004de} that it stated this value for Einstein-Hilbert gravity is $\frac{\eta }{s}=\frac{1 }{4 \pi}$ is preserved for this model. $\frac{\eta }{s}$ is proportional to inverse squared coupling of field theory, $\frac{\eta }{s} \sim \frac{1 }{\lambda^2}$. It means that the coupling of the field theory dual to our model  is the same as the coupling of the field theory dual to the Einstein AdS black brane solution, but the color conductivity is different.

\vspace{1cm}
\noindent {\large {\bf Acknowledgment} } We would like to thank Shahrokh Parvizi, Matteo Baggioli, Mojtaba Shahbazi and Komeil Babaei for useful comments and suggestions. We also thank the referees of CJP for the valuable comments which helped us to improve the manuscript.

\vspace{1cm}
\noindent {\large {\bf Data Availability } } Data generated or analyzed during this study are provided in full within the published article.

\end{document}